\documentclass[aps,eqsecnum,nofootinbib,superscriptaddress,showpacs]
{revtex4}
\usepackage[dvips]{graphicx}  
\usepackage{amsmath,amsfonts,bm}  
\def\be{\begin{eqnarray}}
\def\ee{\end{eqnarray}}
\def\beq{\begin{eqnarray}}
\def\eeq{\end{eqnarray}}

\def\({\left (}
\def\){\right )}
\def\[{\left [}
\def\[{\right ]}

\def\tilr{\tilde{r}}
\bmdefine{\bmx}{\bm{x}}
\bmdefine{\bmz}{\bm{z}}
\newcommand{\tilg}{ \tilde{g} }
\newcommand{\tilG}{ \tilde{G} }
\newcommand{\tilH}{ \tilde{H} }
\newcommand{\tilphi}{ \tilde{\phi} }

\newcommand{\hT}{ \Hat{T} }
\newcommand{\homega}{ \Hat{\omega} }
\newcommand{\deldI}{ \delta^{(d+1)} }
\newcommand{\deld}{ \delta^{(d)} }
\catcode`?=\active \def?{\phantom{0}}
\begin{document}
%
%
%
\title{Quasinormal modes for nonextreme D$p$-branes\\
and\\
thermalizations of super-Yang-Mills theories}
\author{Kengo Maeda}
\email{kmaeda@kobe-kosen.ac.jp}
\affiliation{Department of General Education,
Kobe City College of Technology, Kobe, 651-2194, Japan}
\author{Makoto Natsuume}
\email{makoto.natsuume@kek.jp}
\affiliation{Theory Division, Institute of Particle 
and Nuclear Studies, KEK, High Energy Accelerator Research
Organization, Tsukuba, Ibaraki, 305-0801, Japan}
\author{Takashi Okamura}
\email{okamura@ksc.kwansei.ac.jp}
\affiliation{Department of Physics, Kwansei Gakuin University,
Sanda, 669-1337, Japan}
\date{\today}
\begin{abstract}
The nonextreme D$p$-brane solutions in type II supergravity
(in the near-horizon limit) are expected to be dual
to $(p+1)$-dimensional noncompact supersymmetric Yang-Mills theories
at finite temperature.
We study the translationally invariant perturbations along the branes
in those backgrounds and calculate quasinormal frequencies numerically.
These frequencies should determine the thermalization time scales
in the dual Yang-Mills theories.
\end{abstract}
\maketitle
%
\section{Introduction}
As with any dualities, gauge/gravity dualities or AdS/CFT dualities
(anti-deSitter/conformal field theory dualities)
are interesting in two respects.
are interesting in two respects.
On one side, supergravities give information of dual
Yang-Mills theories in strong coupling regimes such as confinement.

On another side,
Yang-Mills theories give information of supergravities.
Finite temperature gauge/gravity duals are particularly interesting
in this respect.
The original finite temperature gauge/gravity duality
is the duality between ${\cal N}=4$ finite temperature
super-Yang-Mills theory (SYM) and type IIB string theory
in the Schwarzschild-${\rm AdS}_5$
(${\rm SAdS}_5$) $\times S^5$ background \cite{Witten:1998zw}.
Thus, finite temperature gauge/gravity dualities should address
long-standing puzzles in gravity,
such as the singularity problem
\cite{Kraus:2002iv, Hong Liu:2005}~(and references therein),
the information paradox \cite{Maldacena:2001kr,Hawking:2005kf},
and the Gregory-Laflamme instability \cite{GL,Aharony:2004ig}. 

Unfortunately, finite temperature gauge/gravity dualities
have been less studied compared with the zero-temperature AdS/CFT.
First, many evidences of finite temperature gauge/gravity dualities
remain qualitative.
Most well-known evidences are 
\begin{enumerate}
\item The existence of the confinement-deconfinement transition
(the Hawking-Page transition in gravity side \cite{Hawking:1982dh})
\item The large-$N$ dependence of the partition function in each phase
\end{enumerate}
but quantitative understandings are still far.
(The lack of supersymmetry for finite temperature
is obviously the main obstacle.)
Second, backgrounds other than ${\rm SAdS}_5$ have been less studied
compared with the zero-temperature cases.%
\footnote{For zero temperature, many backgrounds
with less supersymmetry are known by adding perturbations
to the ${\cal N}=4$ SYM.
Some examples are the Klebanov-Strassler background and
the Polchinski-Strassler background, which are dual to certain
${\cal N}=1$ SYM \cite{Klebanov:2000hb,Polchinski:2000uf}.}

This paper provides one step toward these directions.
The original AdS/CFT duality is motivated by the near-horizon limit
of the extreme D3-brane.
But the similar dualities are expected for the other D$p$-branes
(if $p<5$) \cite{Itzhaki:1998dd}.
They are expected to dual to $(p+1)$-dimensional SYM.
Some qualitative features have been known for these dualities,
but there are less quantitative studies.
(Various Wilson loops have been computed, {\it e.g.}, see
Ref.~\cite{Brandhuber:1998er} and references therein.
At the zero temperature, the correlation functions are computed
in Ref.~\cite{Sekino:1999av}.) 

Our aim is to compute the quasinormal~(QN) frequencies 
in these backgrounds.
It is an important concept in black hole physics 
and has been widely discussed in the literature.
Moreover, the QN frequencies of an AdS black hole
have an interpretation in the dual gauge theory.
Such a black hole corresponds to a thermal state in the gauge theory.
The QN frequencies measure
how perturbations of black holes decay.
In the dual theory, this corresponds to the process
where a perturbation of the thermal state decays and the system
returns to the thermal equilibrium.
Thus, QN frequencies give the prediction of
the thermalization time scale for the strongly-coupled gauge theory
\cite{HoHu,MossNorman, BK,CKL,Konoplya,ref:MusiriSiopsis}.%
\footnote{As with all dualities, gravity and gauge theories
do not have an overlapping region of validity.
Our results should be regarded as the strong-coupling prediction
of gauge theories (in the region where D-brane description is valid.
See Sect.~\ref{sec:validity}.)}
It has been also argued that these modes govern
the behavior of gauge theory plasmas.
(See, {\it e.g.}, Refs.~\cite{Benincasa:2005iv,Policastro:2002se,
ref:Starinets,ref:NunezStarinets,ref:KovtunStarinets}
and references therein.)

The plan of the present paper is as follows.
First, in the next section, we review gauge/gravity dualities
for D$p$-branes.
In section \ref{sec:EOM}, we present the perturbed equations 
which are translationally invariant along the brane, and briefly 
review a simple numerical method~\cite{HoHu} to obtain QN 
frequencies. 
In section \ref{sec:discussion}, we discuss the numerical results.
We conclude in section \ref{sec:summary} with a summary of our results.
In the Appendix, we briefly 
review QN modes for readers not having sufficient background in them. 
\section{Gauge/gravity dualities for D$p$-branes}\label{sec:review}
\subsection{Bulk geometry}

In this Section,
we quickly review gauge/gravity dualities for D$p$-branes.
The relevant part of type II supergravity action is given by
\be 
\label{original-action}
S=\frac{1}{(2\pi)^7l_s^8} \int d^{10}x\sqrt{-G} \left[~
e^{-2 \phi}~\Bigl(R+4(\nabla\phi)^2\Bigr)
-\frac{1}{2(p+2)!}F_{p+2}^2~\right]. 
\ee
The nonextreme D$p$-branes are written as
\beq 
\label{dp-solution}
ds^2&=&Z_p^{-1/2}(-hdt^2+d\vec{x}_p^2)+Z_p^{1/2}
(h^{-1}dr^2+r^2d\Omega_{8-p}^2), \nonumber \\
g_s^2 e^{-2\phi}&=&Z_p^{\frac{p-3}{2}}, 
\eeq
where $g_s=e^{\phi_\infty}$ and the harmonic functions are given by
\beq 
Z_p(r)&=&1+\left(\frac{r_p}{r}\right)^{7-p}, \qquad 
r_p^{7-p} \sim g_s N l_s^{7-p},\\
h(r)&=&1-\left(\frac{r_0}{r}\right)^{7-p}.
\eeq

Let us take the ``decoupling" limit or the ``near-horizon" limit.
In order to take the limit, we assume $r \sim r_0 \ll r_p$.
Then, $Z_p=1+(r_p/r)^{7-p} \rightarrow (r_p/r)^{7-p}$. 

Introducing a new coordinate $\tilde{r}$, 
\be
\frac{r}{r_p}=\left(\frac{2}{5-p}\right)^{\frac{2}{5-p}}
\left(\frac{\tilde{r}}{r_p}\right)^{\frac{2}{5-p}},  
\ee
the metric becomes
\beq 
ds^2 &\simeq& \left(\frac{2}{5-p} \right)^2
\left(\frac{r}{l} \right)^{\frac{p-3}{2}} 
  \left[-\tilH(\tilde{r})dt^2
+ \left(\frac{\tilde{r}}{l}\right)^2 d\vec{x}_p^2
+\frac{d\tilde{r}^2}{\tilH(\tilde{r})}
+\left(\frac{5-p}{2}l\right)^2d\Omega^2_{8-p}\right],
\label{planarBH}
\eeq 
where $l = r_p$.
$\tilH$ and $\phi$ are rewritten as 
\beq
\tilH(\tilde{r})&=&\left(\frac{\tilde{r}}{l}\right)^2
-\left(\frac{\tilde{r}_0}{l}\right)^{\frac{2(7-p)}{5-p}}
\left(\frac{\tilde{r}}{l}\right)^{-\frac{4}{5-p}}, \nonumber \\ 
g_s^2e^{-2\phi}&=&\left(\frac{5-p}{2}\frac{l}{\tilde{r}}
\right)^{\frac{(7-p)(p-3)}{5-p}}. 
\label{eq:dilaton}
\eeq
It is clear that the metric is conformal
to ${\rm AdS}_{p+2} \times S^{8-p}$ asymptotically if $p<5$.
Solutions with $p \geq 5$ do not have a positive specific heat,
so we consider $p<5$. We shall use such a ``AdS-frame"
to calculate QN frequencies.

As is well-known, the above metric is not the ${\rm SAdS}$ solution
even if $p=3$. The ${\rm SAdS}_5$ solution is given by
\be
ds^2 = - \left( \frac{\tilr^2}{l^2}+1
     -\frac{\tilr_0^4}{l^2\tilr^2} \right) dt^2
+ \left( \frac{\tilr^2}{l^2}+1
-\frac{\tilr_0^4}{l^2\tilr^2} \right)^{-1} d\tilr^2
+\tilr^2 d\Omega_3^2.
\ee
${\rm SAdS}_5$ has the horizon with the topology of $S^p$,
whereas the metric (\ref{planarBH}) has the horizon
with the topology $R^p$.
We call such solutions \lq\lq planar black holes.\rq\rq\,
The $p=3$ planar black hole corresponds to
the large black hole limit of ${\rm SAdS}_5$.
In order to reach the planar black hole from ${\rm SAdS}_5$,
rescale the coordinates
\be
t~\rightarrow~t/\alpha, \qquad\qquad
\tilde{r}~\rightarrow~\alpha\, \tilde{r},
\qquad\qquad {\rm and} \qquad\qquad
\tilde{r}_0~\rightarrow~\alpha\, \tilde{r}_0.
\label{eq:scaling}
\ee
Then, the $S^3$ radius is proportional to $\alpha$,
so $\alpha^2 d\Omega_3^2 \sim \sum d\vec{x}_p{}^2$ for large $\alpha$.
This limit, the planar ${\rm SAdS}_5$ black hole,
is invariant under the above scaling.

We consider such planar black holes from the following reasons.
First, AdS black hole solutions with $S^p$ topology
are not known when the dilaton is nontrivial.
Second, in the SYM description,
SAdS corresponds to a compact SYM on $S^p$,
and a planar black hole corresponds to a noncompact SYM on $R^p$.
The motivation to consider a compact SYM in Ref.~\cite{Witten:1998zw}
is to break the scale invariance (\ref{eq:scaling}).
Without breaking the scale invariance,
one cannot see the confinement/deconfinement transition
in gauge theory.
Here, there is no scale invariance due to the dilaton.
So, it is not clear whether one should consider a compact SYM.%
\footnote{It is not known
if there is a confinement/deconfinement transition in these theories.
The dual geometry seems to suggest that there is no such a transition
in these theories as well;
namely, the specific heat is always positive for (\ref{planarBH}),
so there is no sign of thermal instability.}
\subsection{Validity of supergravity descriptions}\label{sec:validity}
To discuss the validity of supergravity description (\ref{planarBH}),
it is convenient to introduce SYM variables.
The SYM coupling in terms of string variables are
(See, {\it e.g.}, Ref.~\cite{Polchinski:1998rr}) 
\be
g_{YM}^2 =(2\pi)^{p-2}\, g_s\, l_s^{p-3},
\ee
where
$g_{YM}$ is the $(p+1)$-dimensional SYM coupling constant.
The effective (dimensionless) coupling of SYM theories is 
\be
g_{eff}^2 = g_{YM}^2\, N\, \left( \frac{r}{l_s^2} \right)^{p-3}.
\ee

The SYM perturbation theory can be trusted in the region 
\be
g_{eff} \ll 1.
\ee
On the other hand, one can trust supergravity solutions
if both the curvature (in string metric) and the dilaton are small.
Since
\begin{align}
  & e^{\phi} \sim g_{eff}^{(7-p)/2}/N~,
& & l_s^2\, R \sim 1/g_{eff}~,
\end{align}
these conditions imply
\be
1 \ll g_{eff}^2 \ll N^{4/(7-p)}.
\ee
Clearly, the perturbative SYM and supergravity descriptions
do not overlap.
For $p<3$, this gives the following range of $r$
(not the AdS-like coordinate $\tilde{r}$):
\be
(g_{YM}^2 N)^{1/(3-p)} N^{-4/(3-p)(7-p)}
\ll r/l_s^2 \ll (g_{YM}^2 N)^{1/(3-p)}~.
\label{eq:validity}
\ee
(For $p>3$, replace the $\ll$ signs by $\gg$ signs.)
The left-hand side and the right-hand side of these inequalities
come from the dilaton and the curvature, respectively.
They have a diverging dilaton at $r=0$ and a curvature singularity
at $r=\infty$.
One necessary condition to satisfy the above condition is $N \gg 1$.

When the radial coordinate $r$ is outside the region,
different theories (such as M-theory) take over
the type II descriptions.
The radial coordinate has the gauge theory interpretation
as the energy scale.
The phase diagrams are discussed in Ref.~\cite{Itzhaki:1998dd}.
For type IIA-branes, the M-brane description take over at small radius.
As argued later, the M-brane description often reduces to a
SAdS black hole, and there have been extensive works on the subject;
this limit is relatively well-known.
The other limit is the perturbative SYM description at large radius;
again this limit is rather well-known.
Therefore, we focus on the intermediate energy scale
where type II supergravity is a valid description.

In order to calculate QN frequencies,
one places boundary conditions both at the horizon and at infinity.
We henceforth consider the case where the horizon radius lies
inside the region (\ref{eq:validity}).%
\footnote{As a matter of fact, our results are valid even
for black holes with smaller horizon (for $p<3$).
This is because the change of a supergravity description ({\it e.g.},
from the type IIA supergravity to the 11-dimensional supergravity)
does not change the results.
On the other hand, the horizon must be sufficiently smaller than
the right-hand side of Eq.~(\ref{eq:validity}).
This is because the supergravity description must be valid 
not only at the horizon, but also
in the region where QN modes decay. These issues
are discussed in Sect.~\ref{sec:BC}.}
In the large-$N$ limit ($N\rightarrow\infty$ and $g_{YM}\rightarrow0$
with a fixed large $g_{YM}^2 N$),
one can enlarge this region as large as one wishes.
Type II description is not a good description at infinity as well.
One might put a boundary condition at a large radius
but within the region of the validity.
We here assume that such a boundary condition at large radius
does not affect the results significantly. 
We return to the issue of the validity 
and discuss how different boundary conditions may affect our results 
in Sect.~\ref{sec:BC}.
\section{Numerical approaches for QN frequencies} \label{sec:EOM}
\subsection{Basic equations}
After the conformal transformation of the metric (\ref{dp-solution}),
\be
d\tilde{s}^2 = \left(\frac{5-p}{2}\right)^2
  (g_se^{-\phi})^{\frac{1}{7-p}}ds^2
  = \tilG_{MN}\, dx^M\, dx^N, 
\ee
the action~(\ref{original-action}) is transformed as
\beq 
S= \int d^{10}x\sqrt{-\tilde{G}}
\left[~e^{2a\phi} \Big( \tilde{R} +4 b (\tilde{\nabla}\phi)^2 \Big)
-\frac{\gamma^{2(p+1)}e^{-2a\phi}}{2(p+2)!}\tilde{F}_{p+2}^2~\right],
\eeq
up to a constant, where 
\be 
\label{b-coefficient}
a=\frac{p-3}{7-p}, \qquad b=\frac{(p-1)(p-4)}{(7-p)^2},
\qquad \gamma=\frac{5-p}{2}~g_s^{\frac{1}{7-p}}.
\ee
 
Since we are not interested in the perturbation on $S^{8-p}$ sphere, 
we assume an ansatz for the metric: 
\be 
\label{8-p sphere}
d\tilde{s}^2 = \underline{g}_{\mu\nu}(x^\rho)dx^\mu dx^\nu
+\left( \frac{5-p}{2}\, l \right)^2 d\Omega^2_{8-p},
\ee
where $\Lambda=-\frac{1}{l^2}\frac{(9-p)(7-p)}{(5-p)^2}$
and greek indices run from $0$ to $p+1$. 
After the compactification, one gets a $(p+2)$-dimensional action
\cite{Townsend}
\beq 
\label{p+2 action}
S&=& \int d^{p+2}x~\sqrt{-\underline{g}}~e^{2a\phi}~
\Big( \underline{R}-2\Lambda + 4 b\, (\underline{\nabla}\phi)^2 \Big)~,
\eeq
up to a constant.
The $p=0$ case reduces to two-dimensional gravity coupled to a scalar,
so the system locally has no dynamical degrees of freedom.
Hereafter, we focus on $1 \le p \le 4$.

For simplicity, we shall only consider the perturbations
which are translationally invariant along the brane.
Then, one can set the metric as
\begin{align}
  & d\underline{s}_{(p+2)}^2 = \underline{g}_{\mu\nu}\, dx^\mu dx^\nu
  = {}^{(2)}\!g_{ab}(x^c)\, dx^a dx^b
  + e^{2 \zeta(x^a)}\, d\vec{x}_p{}^2~,
\end{align}
where $x^a=(t, \tilr)$.
And the action (\ref{p+2 action}) becomes
\begin{align}
  & S = \int d^2 x~\sqrt{ - {}^{(2)}\!g }~e^{\psi_2}
  \left[~{}^{(2)}\!R - 2\, \Lambda
  - \frac{p\, (p-3)^2}{9-p}\, \left( {}^{(2)}\!\nabla \psi_1 \right)^2
  + \frac{4}{9-p} \left( {}^{(2)}\!\nabla \psi_2 \right)^2~\right]~,
\label{eq:dim_redunction}
\end{align}
where
\begin{align}
  & \psi_1 = \frac{5-p}{(7-p)(p-3)}~2\, \phi - \zeta~,
& & \psi_2 = 2 a\, \phi + p\, \zeta~,
\end{align}
The system is two-dimensional gravity
coupled to two scalars, $\psi_1$ and $\psi_2$,
and there is only one dynamical degrees of freedom.
We need to find the dynamical degrees of freedom and
obtain its perturbative equation around the background solutions
given by Eqs.(\ref{planarBH}) and (\ref{eq:dilaton}).
Fortunately, since the background solution for $\psi_1$ is constant,
it is easy to show that its perturbation $\delta \psi_1$
is gauge invarint at the perturbative level
\cite{ref:KodamaSasaki},
and the equation of motion is given by%
\footnote{This equation is equivalent to the massless scalar field
minimally coupled to ten-dimensional Einstein metric.
}
\begin{align}
  & 0 = \partial_a \left[~\sqrt{ - {}^{(2)}\!\bm{g} }~
    {}^{(2)}\!\bm{g}^{ab}~e^{\bm{\psi}_2}~
    \partial_b\, \delta \psi_1~\right]~,
\label{eq:perturbed-EOM}
\end{align}
where the bold face letters are background quantities.

One can show that this equation is invariant
under the scaling (\ref{eq:scaling}) with $l$ fixed.
By the same argument as in Ref.~\cite{HoHu},
this scale invariance means that QN frequencies are proportional to
$\tilr_0$, or the black hole temperature $T \propto \tilr_0/l^2$.
Hence, it is convenient to introduce dimensionless coordinates
$\tau=\tilr_0\, t/l^2$ and $\eta=\tilr/\tilr_0$, and
\begin{align}
  & Z = e^{ \bm{\psi}_2/2 }~\delta \psi_1~.
\end{align}
Then, Eq.(\ref{eq:perturbed-EOM}) is rewritten by
\beq
&& -\partial_{\tau}^2 Z
   = \left( - \partial_{\eta_*}^2  + V\, \right)\, Z
   = \big[~- H \partial_\eta (H \partial_\eta)  + V~\big]\, Z~,
\label{eq:noremalized_EOM}
\eeq 
where
%
\begin{align}
  & H = \eta^2 (1-\eta^{-q})~,
& & V = \nu\, H~\big( \nu+1 + \nu\, \eta^{-q} \big)~,
\\
  & \nu = \frac{9-p}{2(5-p)}~,
& & q = \frac{2(7-p)}{5-p}~,
\end{align}
and the tortoise coordinate $\eta_*$ is defined by 
\be 
\eta_*=\int d\eta/H~.
\ee
The potential $V$ is monotonically increasing function
of $\eta$ and $V\, \vert_{\eta=1} = 0$,
so the potential $V$ is positive-definite outside the horizon
($\eta > 1$).
\subsection{A numerical method to obtain QN frequencies}
Following Horowitz and Hubney's method~\cite{HoHu},
let us calculate the QN frequencies for D$1$, D$2$, and D$4$-branes.%
\footnote{For $p=3$,
the QN frequencies for the planar black hole correspond 
to the frequencies for the large black hole limit of ${\rm SAdS}_5$,
which were calculated in Refs.~\cite{HoHu,ref:Starinets}. }
For the later convenience, we shall define 
$\psi$ as 
\be 
\label{inner-boundary}
Z = e^{-i \hat{\omega} \tau}
  (1- \eta^{-q})^{-i \hat\omega/q}~\psi(\eta)~,
\ee
and introduce a new coordinate $x=\eta^{- 2 \sigma/(5-p)}$,
where $\sigma = 1$ for $p=$ even and $\sigma = 2$ for $p=$ odd.
When $p<5$, the horizon and the infinity
correspond to $x=1$ and $x=0$, respectively.
Then, one gets a Fuchs-type differential equation:  
\begin{align} 
  & s(x) \frac{d^2}{dx^2}\, \psi(x)
  + \frac{t(x)}{x-1}\, \frac{d}{dx}\, \psi(x)
  + \frac{u(x)}{(x-1)^2}\, \psi(x) = 0~,
\label{eq:Fuchs-Eq}
\end{align} 
where
\begin{align} 
  & s(x) = \left( x\, \frac{ x^{(7-p)/\sigma} - 1 }{x - 1} \right)^2~,
\\
  & t(x) = \frac{x}{2 \sigma}\, \frac{ x^{(7-p)/\sigma} - 1 }{x-1}~
   \Big[~5 - 2 \sigma - p
   + \big\{ 9 + 2 \sigma - p - 2i (5-p)\, \hat\omega \big\}\, 
   x^{(7-p)/\sigma}~\Big]~,
\\
  & u(x) = - \left( \frac{5-p}{2 \sigma} \right)^2\,
  \Big[~( {\hat\omega} + i\, \nu )^2\, x^{2(7-p)/\sigma}
    - ( 1 + 2 i\, \nu\, \hat{\omega} )\, x^{(7-p)/\sigma}
  - \hat{\omega}^2\, x^{(5-p)/\sigma} + \nu (\nu+1) \Big]~.
\end{align} 

We solve these equations by expanding $\psi$
around the horizon~($x=1$). 
For the power expansion of $\psi$ about $x=1$ to be applicable
up to the asymptotic infinity ($x=0$),
the radius of convergence must reach $x=0$. 
Let us examine the singularity structure of Eq.(\ref{eq:Fuchs-Eq})
on the complex $x$-plane.
Equation~(\ref{eq:Fuchs-Eq}) has regular singular points when 
$x=0$ and $x^{(7-p)/\sigma} = 1$, 
namely 
$x=e^{2\pi i\, \frac{m}{7-p}}\,(m=1, 2, \cdots, 7-p)$ 
for $p=$ even
and $x=e^{2 \pi i\, \frac{2m}{7-p}}\,(m=1, 2, \cdots, \frac{7-p}{2})$
for $p=$ odd. 
So, the nearest singular point of $x=1$ is $x=0$
and the power expansion of $\psi$ about $x=1$ is applicable
up to the asymptotic infinity ($x=0$).

The QN modes are the solutions of Eq.(\ref{eq:Fuchs-Eq})
with the following boundary conditions: 
\begin{enumerate}
\item[(i)] The scalar wave is purely ingoing near the horizon,
$Z \sim e^{- i \homega (\tau + \eta_*) }$.
\item[(ii)] The scalar wave decays at infinity,
$Z \sim e^{-i \homega \tau} \eta^{-(\nu+1)}$.~
(Another mode diverges.)
\end{enumerate}
The purely ingoing mode 
is expressed near the horizon as 
\be 
Z\sim e^{-i\hat{\omega} \tau}e^{-i\hat{\omega} \eta_*}
=e^{-i\hat{\omega} \tau}e^{-i\hat{\omega}\int d \eta/H}
\sim e^{-i\hat{\omega} \tau}(\eta-1)^{-i\hat\omega/q}
\sim e^{-i\hat{\omega} \tau}
(1-\eta^{-q})^{-i\hat{\omega}/q},  
\ee
which is just the same prefactor in front of $\psi$
in Eq.~(\ref{inner-boundary}).
So, the solution satisfying the condition (i) has the form
\be 
\psi(x)=\sum^\infty_{n=0}a_n(x-1)^n, 
\ee
where $a_0$ is a non-zero constant. 
Then, the coefficients $a_n$ are obtained
by the following recursion relation: 
\be 
\label{recursion}
a_n = -\frac{1}{ n(n-1)\, s_0 + n\, t_0 }~
\sum^{n-1}_{k=0}~[~k(k-1)\, s_{n-k} + k\, t_{n-k} + u_{n-k}~]\, a_k~,
\ee
where $s_n$, $t_n$, and $u_n$ are the $n$-th order coefficients
of the expansions around $x=1$,
{\it e.g.}, $s(x)=\sum_{n=1}^\infty s_n\, (x-1)^n$.
Since the equations are linear,
the coefficient $a_0$ is a free parameter (we set $a_0=1$).

In order to find the QN frequencies,
we need to find the solution satisfying
the latter boundary condition (ii):
\be 
\label{suminfinity}
\psi(0)=\sum^\infty_{n=0}a_n(-1)^n=0~,
\ee 
which gives a polynomial equation of $\hat{\omega}$.

Before we solve the above equation of $\homega$
for $p=1,2$, and $4$ numerically,
we comment on some general properties of QN frequencies:
\begin{itemize}
\item QN frequencies are symmetrically distributed
  with respect to the imaginary axis of the complex $\homega$-plane,
  {\it i.e.}, there is a symmetry
  $\homega~\leftrightarrow\, - \homega^*$.
\item The imaginary part of the QN frequency is negative
  so that the system in our interests is stable.
\end{itemize}

The former statement is proved as follows:
Consider any QNM, $Z_{\homega}$ with a QN frequency $\homega$.
Then, $\big( Z_{\homega} \big)^*$ is also a solution of
Eq.(\ref{eq:noremalized_EOM}) with the frequency
$\homega \rightarrow - \homega^*$.
Furthermore, its asymptotic form is
$\big( Z_{\homega} \big)^* \sim e^{-i (-\homega^*) (\tau + \eta_*) }$
near the horizon, and $\big( Z_{\homega} \big)^*
\sim e^{-i (-\homega^*) \tau }\, \eta^{-(\nu+1)}$
near the infinity.
This means that $\big( Z_{\homega} \big)^*$ is also a QNM whose
QN frequency is $-\homega^*$.

As usual,
the latter statement is proved by the \lq\lq energy integral\rq\rq%
\cite{HoHu},
thanks to the positivity of the potential $V$ outside the horizon,
$\eta>1$.
\section{Discussion of results} \label{sec:discussion}
To solve Eq.~(\ref{suminfinity}) numerically, we find  zeros
of a partial sum $\psi_N=\sum^N_{n=0}a_n(-1)^n$ for a large $N$
using MATHEMATICA.
To obtain an accurate value of QN frequencies,
we need to compute on the order of $N=100$.
\footnote{For example, for $p=2$,
the $N=90$ results differ from the $N=100$ results by about $10^{-5}$.
As a check, we have also applied the numerically
stable continued fraction method
presented by Leaver~\cite{Leaver:1990}
and checked that the numerical values coincide with the ones
obtained by MATHEMATICA with a good accuracy.}

\begin{figure*}[htb]
  \begin{center}
     \includegraphics[width=12.0cm,clip]{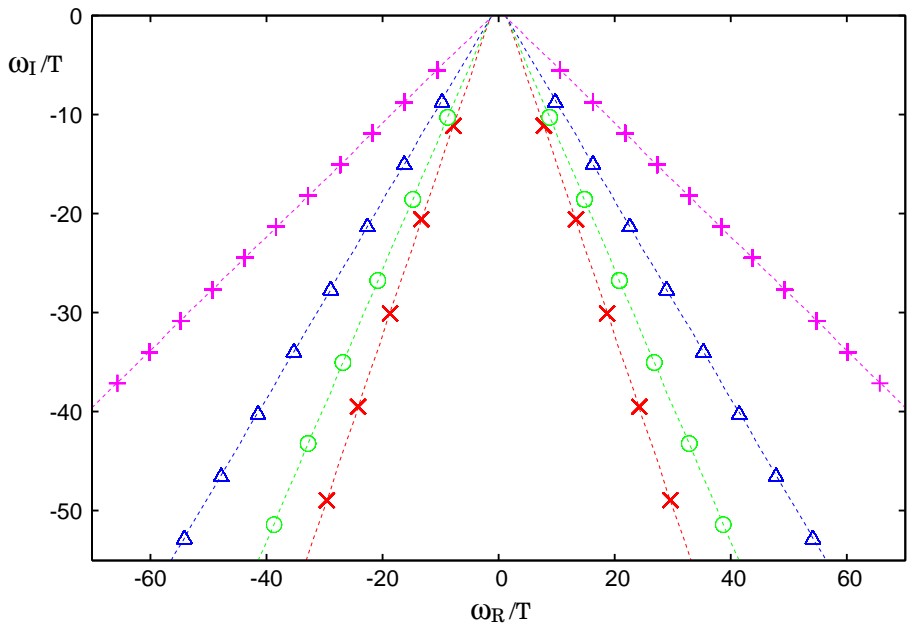}
  \end{center}
\caption{\label{fig:QNM}
QN frequencies for D$1$($\times$), D$2$({\large $\circ$}),
D$3$({\small $\triangle$}) and D$4$-branes({\small $+$}).
Each straight line is determined by the least-square fitting.}
\end{figure*}

We present the QN frequencies for $p=1$, $p=2$, and $p=4$
cases in Table~\ref{tab:QNM},
and their distribution on the complex $\homega$-plane
in Fig.~\ref{fig:QNM}.
The numerical values of QN frequencies are normalized by
the dimensionless Hawking temperature
$\hT=(1/4\pi) dH/d\eta|_{\eta=1}=q/4\pi$.
%
\begin{table}[htb]
\begin{center}
\begin{tabular*}{\textwidth}{@{\extracolsep{\fill}}ccccccc}
\hline\hline
      \multicolumn{1}{c}{~}
    & \multicolumn{2}{c}{$p=1$}
    & \multicolumn{2}{c}{$p=2$}
    & \multicolumn{2}{c}{$p=4$}
 \\
 Mode & $\hat{\omega}_R/\hT$ & $\hat{\omega}_I/\hT$
      & $\hat{\omega}_R/\hT$ & $\hat{\omega}_I/\hT$
      & $\hat{\omega}_R/\hT$ & $\hat{\omega}_I/\hT$
 \\ \hline
 $0$ & ?7.747 & -11.158
     & ?8.710 & -10.260
     & 10.488 & ?-5.472
 \\
 $1$ & 13.242 & -20.594
     & 14.775 & -18.509
     & 16.232 & ?-8.709
 \\
 $2$ & 18.700 & -30.023
     & 20.780 & -26.743
     & 21.805 & -11.889
 \\
 $3$ & 24.149 & -39.450
     & 26.770 & -34.972
     & 27.319 & -15.051
 \\ \hline\hline
\end{tabular*}
\end{center}
\caption{QN frequencies for $p=1$, $p=2$, and $p=4$. 
$\hat{\omega}_R$ and $\hat{\omega}_I$ are
real and imaginary parts of $\hat{\omega}$, respectively. }
\label{tab:QNM}
\end{table}%
%
%
%
As is apparent from Fig.\ref{fig:QNM},
the QN frequencies for each $p$ are distributed on the straight line.
The QN frequencies are approximately given
by the modes $n$~($n=0, 1, 2, \cdots$) of the formula below:
\begin{equation}
  \homega_n/\hT = \alpha_p~n + \beta_p~,
\end{equation}
where
\begin{align}
  & \alpha_p
  = \begin{cases}
      5.45 - 9.43\, i & \text{for~~} p=1 \cr
      6.00 - 8.23\, i & \text{for~~} p=2 \cr
      6.32 - 6.29\, i & \text{for~~} p=3 \cr
      5.52 - 3.16\, i & \text{for~~} p=4 \cr
  \end{cases}~,
& & \beta_p
  = \begin{cases}
      7.80 - 11.17\, i & \text{for~~} p=1 \cr
      8.76 - 10.27\, i & \text{for~~} p=2 \cr
      9.88 -  8.66\, i & \text{for~~} p=3 \cr
     10.66 -  5.53\, i & \text{for~~} p=4 \cr
  \end{cases}~.
\end{align}
This property that the QN frequencies are approximately evenly spaced
with $n$ is numerically observed
for the scalar, vector, and gravitational perturbations
on the ${\rm SAdS}_4$ black holes \cite{CKL}
and on the ${\rm SAdS}_5$ black holes
\cite{ref:Starinets,ref:NunezStarinets,ref:KovtunStarinets}.
The property is analytically shown
for a minimally coupled massive scalar \cite{ref:MusiriSiopsis}
and the vector perturbations \cite{ref:NunezStarinets}
on the ${\rm SAdS}_5$ black hole.
For highly-overtone QN frequencies, one can make a comprehensive research
analytically \cite{ref:NatarioSchiappa}.
However, its origin and significance are far from obvious.%
\subsection{$p=1$ and $p=4$}
For both $p=1$ and $p=4$, the dilaton gravity 
action~(\ref{p+2 action}) is simply written by 
\be
\label{p+2-simple}
S\sim\int d^{p+2}x~\sqrt{-\underline{g}}~e^{2a\phi}~
\Big( \underline{R} - 2 \Lambda \Big)~.
\ee
So, under the metric ansatz: 
\be 
\label{p+3-metric}
ds_{(p+3)}^2=\underline{ds}^2+\left(\frac{5-p}{2}\right)^2
\left(\frac{e^\phi}{g_s}\right)^{\frac{4(5-p)}{(7-p)(p-3)}}dz^2
\ee
one can embed the $(p+2)$-dimensional action~(\ref{p+2-simple})
into the $(p+3)$-dimensional action: 
\be 
S\sim \int d^{p+3}x\sqrt{-g_{(p+3)}}~
 \Big(~{}^{(p+3)}\! R - 2 \Lambda~\Big)~. 
\ee
For $p=4$, the embedding can be understood as 
the M-theory embedding (M5-brane) of the D4-brane, 
and this M5-brane reduces to the planar ${\rm SAdS}_7$ black hole 
(in the near-horizon limit). 

For $p=1$, D1-brane does not have a M-theory 
embedding because it is a type IIB object. 
However, various dualities relate the D1-brane to the M2-brane.
Under the four-dimensional 
pure gravity theory, the metric~(\ref{p+3-metric}) becomes   
\be 
ds^2=-(\tilde{r}^2-\tilde{r_0}^3/\tilde{r})dt^2
+(\tilde{r}^2-\tilde{r_0}^3/\tilde{r})^{-1}dr^2+
\tilde{r}^2(dx^2+dz^2), 
\ee
where we set $l=1$ for simplicity.
This metric corresponds to the planar ${\rm SAdS}_4$ metric.
So, one can compare our $p=1$ results
with the ones for the ${\rm SAdS}_4$ in the large black hole limit.
Table~\ref{tab:comparison} is the comparison
with the results by Cardoso, et.al. \cite{CKL}
corresponding to the $\ell=2$ gravitational perturbations (even parity)
of a large ${\rm SAdS}_4$ black hole ($l=1$ and $\tilr_0=100$).
Their results are normalized with respect to
the Hawking temperature $T = ( q/4\pi )\, (\tilde{r}_0/l^2)$.
One can easily see that our results agree well
with the ${\rm SAdS}_4$ results in the large black hole limit.
%
\begin{table}[htb]
\begin{center}
\begin{tabular*}{12cm}{@{\extracolsep{\fill}}ccccc}
\hline\hline
      \multicolumn{1}{c}{~}
    & \multicolumn{2}{c}{Ours}
    & \multicolumn{2}{c}{Cardoso, et.al}
 \\
 Mode & $\homega_R/\hT$ & $\homega_I/\hT$
      & $\omega_R/T$ & $\omega_I/T$
 \\ \hline
 $0$ & ?7.747 & -11.158
     & ?7.748 & -11.157
 \\
 $1$ & 13.242 & -20.594
     & 13.244 & -20.591
 \\
 $2$ & 18.700 & -30.023
     & 18.703 & -30.020
 \\
 $3$ & 24.149 & -39.450
     & 24.153 & -39.446
 \\ \hline\hline
\end{tabular*}
\end{center}
\caption{QN frequencies corresponding to $\ell=2$ gravitational
perturbations (even parity) of a large Schwarzschild-AdS black hole
with $\tilr_0=100$ and $l=1$.}
\label{tab:comparison}
\end{table}%
%
%
\subsection{$p=2$}

On the other hand, the $p=2$ case is not related to
a higher-dimensional SAdS, contrary to the $p=1$ and $p=4$ cases. 
Obviously, one can always embed type IIA objects into M-theory.
However, the resulting geometries are not SAdS black holes,
and QN frequencies for such geometries are unknown.

The D2-brane is embedded as a M2-brane, but the 
embedding of the geometry (\ref{planarBH}) is not the ${\rm SAdS}_4$.
This is because the embedding corresponds not to the standard M2-brane,
but rather corresponds to the so-called ``smeared M2-brane."
In a sense, we have obtained QN frequencies of
the ``smeared M2-brane" via the D2-brane. 

Even the smeared M2-brane description is not valid at lower energy.
The smeared M2-brane becomes unstable at lower energy
due to the Gregory-Laflamme instability
and decays into the M2-brane on a circle.
Then, the ${\rm SAdS}_4$ calculation suffices
for such a small black hole \cite{Itzhaki:1998dd}. 
\subsection{Sensibility on the boundary condition}\label{sec:BC}
We placed the Dirichlet condition at infinity
to compute QN frequencies.
However, supergravity description often breaks down at infinity.
So, strictly speaking, one must place an ultraviolet cutoff
and put a boundary condition at the large finite radius.
Here, we discuss how such a boundary condition may change our results.

First of all, it is not clear what boundary condition one must impose.
The out-going wave certainly leaks out to the asymptotic infinity.
So, a simple Dirichlet condition at the cutoff does not suffice.
However, appropriate boundary condition is not clear,
so here we use a Dirichlet boundary condition for illustration
to see if our results are sensitive to the boundary condition.

\begin{figure}[htb]
  \begin{center}
    \includegraphics[width=7.3cm,height=5.0cm,clip]{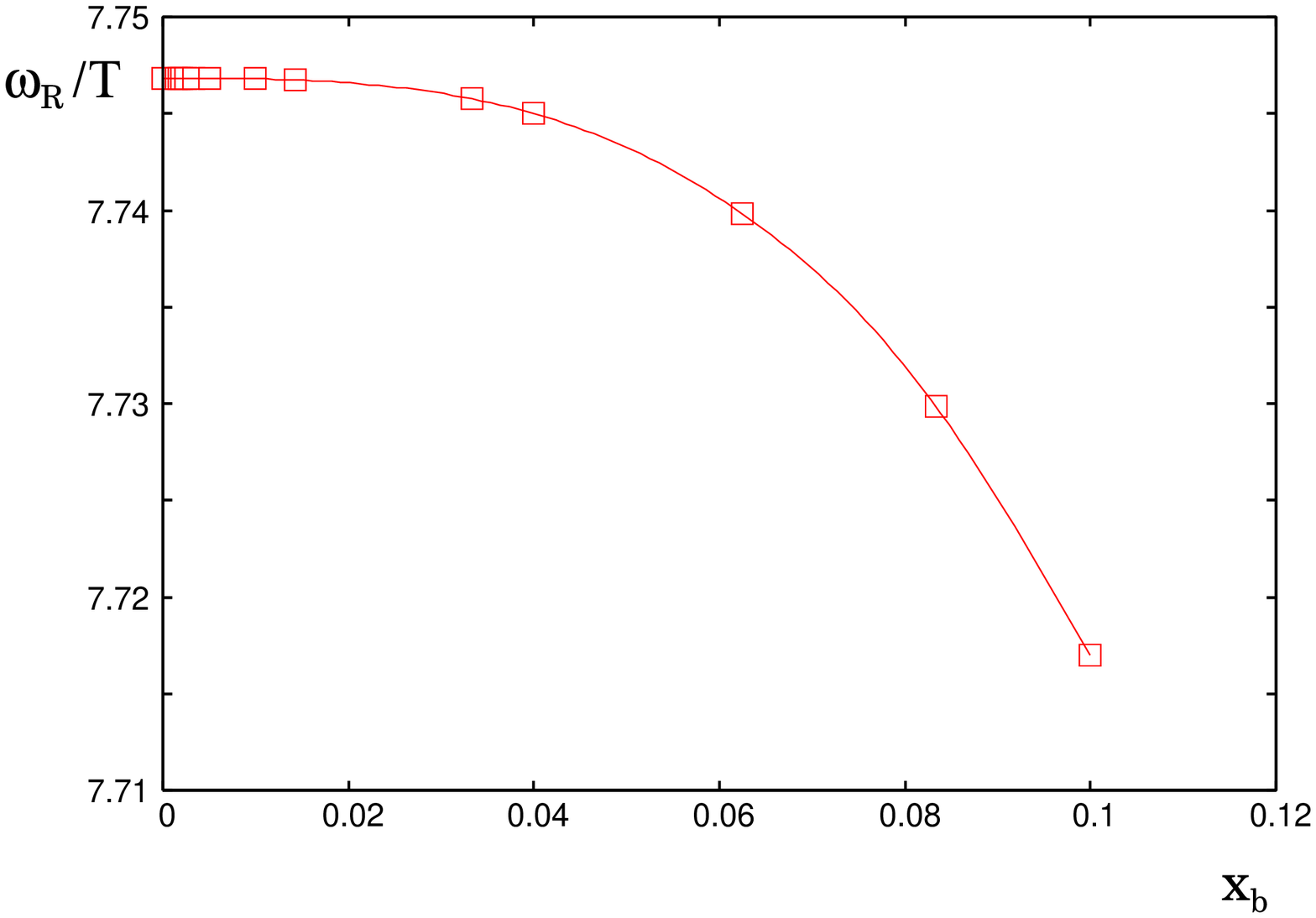}
    \includegraphics[width=7.3cm,height=5.0cm,clip]{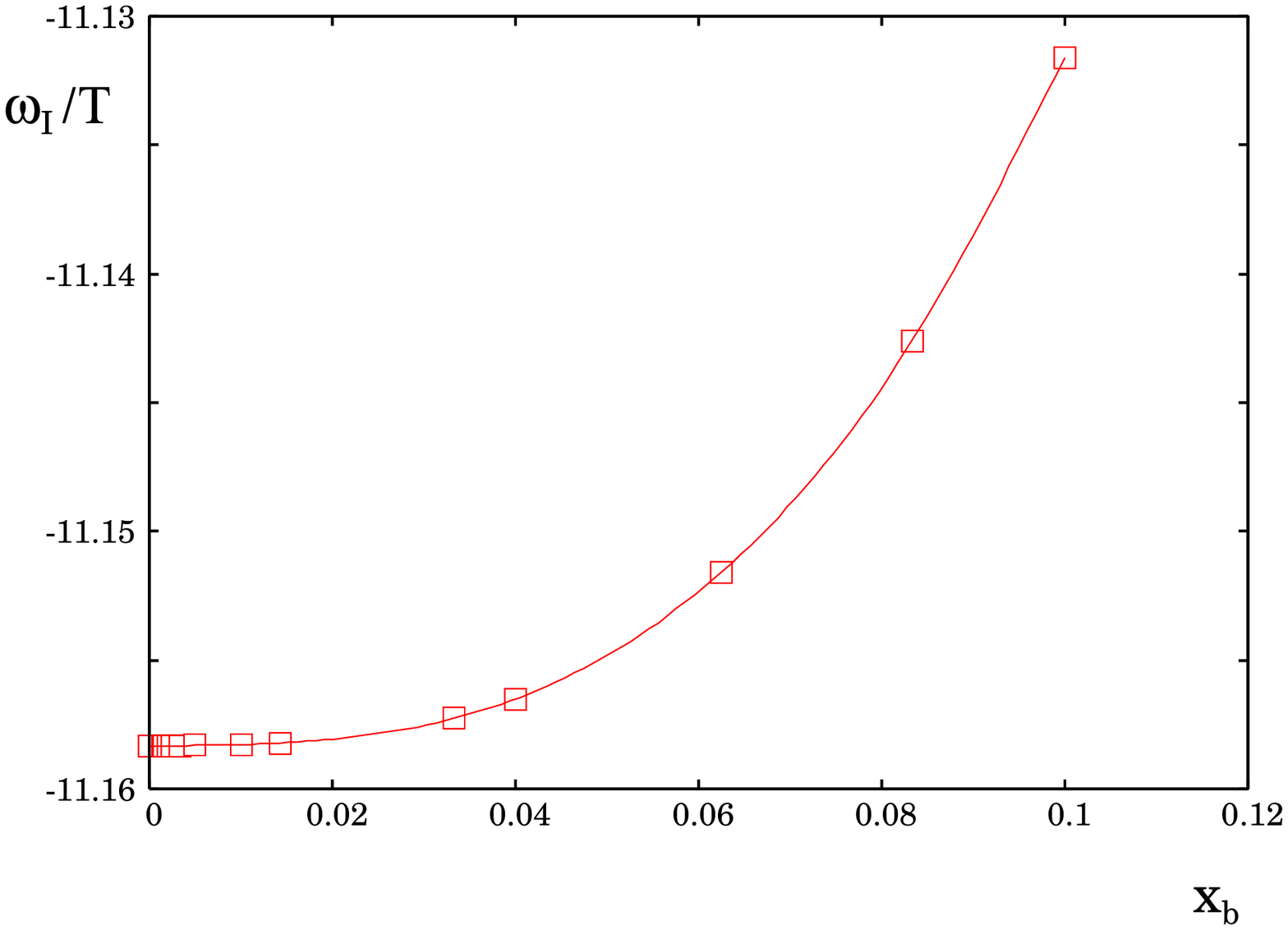}
  \end{center}
\caption{\label{fig:QNM_epsilon}
The dependence of the lowest QN frequency
on the location of the boundary condition $x_b$.
The left figure (right figure) is the real (imaginary) part of
the QN frequency. }
\end{figure}

Figure~\ref{fig:QNM_epsilon} is the result of QN frequencies
(for $p=1$) by imposing the Dirichlet condition
at various radii $x_b=(r_0/r_b)^2$.
(Here, $r_b$ is the location of the boundary condition
in the original D-brane coordinate $r$.)
As the figure shows, the result begins to converge
for a sufficiently large boundary radius
(say, $x_b<0.02$ or $r_b>r_0/\sqrt{0.02}$).
For the boundary condition placed in the plateau region,
the result is effectively the same as the one
with the boundary condition at infinity.
In other words, the result is insensitive to the boundary condition.
%
%

Let us reinterpret the result in terms of SYM variables.
The D1 description is valid
for $g_{YM}N^{1/6}\ll r/l_s^2 \ll(g_{YM}^2 N)^{1/2}$ in SYM variables.
Actually, gravity description is valid even inside the infrared cutoff.
The type IIB fundamental string description takes over
(via type IIB S-duality),
but the gravity computation is the same as the D1 case.
Supergravity descriptions are valid as long as
$g_{YM}\ll r/l_s^2 \ll(g_{YM}^2 N)^{1/2}$.

The theory has two independent parameters $g_{YM}$ and $N$.
The infrared cutoff can be controlled by $g_{YM}$
whereas the ultraviolet cutoff can be controlled by $N$.
Thus, the condition $x_b<0.02$ must be satisfied
for a large enough $N$. This is indeed true. 

In order for the QN frequency to be insensitive
to the boundary condition, $r_b$ must lie within this region.
This implies $g_{YM}\ll r_0/l_s^2 \ll (0.02 g_{YM}^2 N)^{1/2}$.
This condition has a solution if $N \gg 50$.%
\footnote{If one chooses a boundary condition different
from the Dirichlet boundary condition,
the conclusion could be totally changed
(even the existence of the plateau region).
So, one should not take the conclusion seriously.
Our point here is that there exists at least one boundary condition,
where the Dirichlet condition at infinity gives a sufficient accuracy.}
\section{Summary} \label{sec:summary}
We have computed the scalar QN frequencies
for asymptotically AdS black holes (in appropriate frame),
which correspond to the decoupling limit of nonextreme D$p$-branes
(for $p<5$).
We consider the translationally invariant perturbations
along the branes.
The dual gauge theory is described by $(p+1)$-dimensional
super-Yang-Mills theories at finite temperature. 

As discussed in Sect.~\ref{sec:discussion},
one can embed the system into higher dimensions for $p=1$ and $p=4$.
Then, the bulk geometries become just SAdS black holes.
Thus, these cases can be reproduced by QN frequency calculations
of standard SAdS black holes.
(However, they correspond to different region of the validity,
so one must be careful to its interpretation.
For example, the $p=4$ case reduces to the M5-brane,
but this must be interpreted as the recovery of conformality
at high energy.)
This gives a nice check of our approach since our perturbation
is more involved compared with the standard cases.
(Only the minimally-coupled test scalar field is often considered
to calculate QN frequencies.)
We explicitly checked that our results coincide with the SAdS results.

On the other hand, $p=2$ case is not obtained
from higher dimensional SAdS black hole.
Thus, the embedding of the $p=2$ case
does not help to simplify the calculation.%
\footnote{This fact also applies to the $p=0$ 
case. The D0-brane is a M-theory Kaluza-Klein state, so this 
corresponds to the near-horizon limit of a $pp$-wave.}
The fact that QN frequencies are evenly spaced had been observed for
pure gravity black holes.
Our $p=2$ result may indicate that this is
true even for some dilatonic AdS black holes.
%
%

It is difficult to calculate the thermalization time scale
in the gauge theory.
But Ref.~\cite{HoHu} pointed out that the time scale is
likely to be independent of the 't~Hooft coupling.
A free field theory never thermalizes
and at weak coupling the time scale should be very long.
So, clearly the time scale depends on the coupling at weak coupling.
However, in the strong coupling,
the relaxation time scale is expected
to be the order of the thermal wavelength.

In the dual supergravity description, this means that
the QN frequencies depend only on the Hawking temperature. 
This expectation is indeed true for SAdS black holes \cite{HoHu},
and we found that it is also true for D$p$-branes.
(Table~\ref{tab:QNM} shows that QN frequencies are linear
in temperature.)

Finite temperature gauge/gravity dualities deserve further study.
To qualitatively check the duality,
one wishes to check the time scale in the dual theory.
Currently, it is difficult to calculate the time scale
in gauge theories.
Such a calculation has not been carried out even in the SAdS cases.
But gauge theory understanding is essential
to solve the long-standing puzzles such as the singularity problem,
the information paradox, and the Gregory-Laflamme instability.

\vspace*{0.3cm}
{\bf Note added}:
After this work was completed, we were informed about work by Iizuka,
Kabat, Lifschytz, and Lowe \cite{Iizuka:2003ad},
where they also computed QN frequencies for nonextreme D$p$-branes.
The differences from our paper are as follows:
(1) We computed not only the lowest modes,
but also higher modes as well. As a result, we are able to see
how eigenvalues are distributed in the complex plane.
(2) The perturbation considered in the paper eventually becomes the same as ours.
However, they regarded the perturbation simply as
a minimally-coupled test scalar field,
whereas we derived the equation from a combination of the gravitational
and dilaton perturbations, which is part of type II spectrum
in those backgrounds.
(3) We also consider the validity of supergravity descriptions.

We would like to thank D. Kabat for communicating their results.

\begin{acknowledgments}
We would like to thank V. Hubeny, Y. Sekino, and T. Yoneya
for discussions.
The research of M.N.\ was supported in part by
the Grant-in-Aid for Scientific Research (13135224)
from the Ministry of Education, Culture, Sports, Science
and Technology, Japan.
\end{acknowledgments}
\appendix*
\section{Review of Quasinormal Modes}
We briefly review quasinormal modes
(for a comprehensive review, see \cite{ref:KokkotasSchmidt}).

Let us consider the initial-value problem of
a linear wave equation for a scalar field $\Phi$
\begin{align}
  & \big( - \square + V(x)~\big)\, \Phi(x) = 0~.
\label{eq:append-EOM}
\end{align}
As is well-known, we can formally solve the initial-value problem
using the retarded Green function $G_R$ as
\cite{ref:MorseFeshbach}
\begin{align}
  & \Phi(x)
  = \int_{\Sigma_0} d\Sigma(z)~\Big[~
    \Big( n^\mu \partial^z_\mu\, G_R(x\,; z) \Big)~\Phi(z)
  - G_R(x\,; z)~\Big( n^\mu \partial^z_\mu\, \Phi(z) \Big)~\Big]~,
\label{eq:append-formal_solution}
\end{align}
where $d\Sigma(z)$ is the infinitesimal surface element
of the initial surface $\Sigma_0$
and $n^\mu$ is future-directed unit normal vector to $\Sigma_0$.

The retarded Green function $G_R(x\,; z)$
in $(d+1)$-dimensional spacetime
is the solution of the inhomogeneous wave equation
\begin{align}
  & \big( - \square_x + V(x)~\big)\, G_R(x\,; z)
  = - \deldI(x-z)/\sqrt{-g(x)}~,
\label{eq:def-Green_eqn}
\end{align}
satisfying the causal condition,
$G_R(x\,; z) = 0$ for $x \notin J^+(z)$.

For static spacetimes, $ds^2
= N^2(\bmx)\, \big( - dt^2 + \gamma_{ij}(\bmx) dx^i dx^j\, \big)$,
so the Green function is time-translationally invariant,
$G_R(x\,; z) =: G_R(t, \bmx\,; \tau, \bmz)
= G_R(t - \tau, \bmx\,; 0, \bmz)$, and
it is convenient to use the \lq\lq frequency-domain Green function%
\rq\rq\, $\tilg_R$ defined by
\begin{align}
  & \tilg_R(\bmx\,; \bmz|s)
  = \left[~N(\bmx)~N(\bmz)~\right]^{(d-1)/2}~
  \int^\infty_{0^-} dt~e^{-s\, t}~G_R(t, \bmx\,; 0, \bmz)~,
\label{eq:append-L_trans}
\end{align}
which is simply the Laplace transform of $G_R$ (with
the weight $N^{(d-1)/2}$ inserted for convenience).
If there exists such a Laplace transform $\tilg_R$,
an abscissa of convergence $\sigma$ exists and
$\tilg_R$ is well-defined for $\Re(s) > \sigma$.
Then, $\tilg_R$ satisfies the equation,
\begin{align}
  & \left[~s^2 - \Delta_{\bmx} 
  + \cdots~\right]\, \tilg_R(\bmx\,; \bmz | s)
  = - \deld(\bmx-\bmz)/\sqrt{\gamma(\bmx)}~,
\label{eq:tilg_eqn}
\end{align}
where $\Delta$ is the Laplacian with respect to
the metric $\gamma_{ij}$ and the dots denote the terms by the 
\lq\lq effective potential\rq\rq.
After solving Eq.(\ref{eq:tilg_eqn})
with a suitable boundary condition, $G_R$ is obtained by
\begin{align}
  & G_R(t, \bmx\,; 0, \bmz)
  = \left[~N(\bmx)~N(\bmz)~\right]^{(1-d)/2}~
  \int^{\sigma + i\, \infty}_{\sigma - i\, \infty}
     \frac{ds}{2 \pi\, i}~e^{s\, t}~
     \tilg_R(\bmx\,; \bmz|s)~.
\label{eq:append-inv_L_trans}
\end{align}

One can define $\tilg_R$ for $\Re(s) < \sigma$
by the analytical continuation from $\Re(s) > \sigma$.
Provided that
$|s\, e^{s t}\, \tilg_R| \xrightarrow{|s| \rightarrow \infty} 0$
for $\Re(s) < \sigma$,
the contribution to $G_R$
in Eq.(\ref{eq:append-inv_L_trans}) comes from
the singularities of $\tilg_R$ in $\Re(s) < \sigma$.
The pole singularities of $\tilg_R$, $\{s_n\}_{n=0,1,2, \cdots}$,
contribute to $G_R$ as
\begin{align}
  & G_R(t, \bmx\,; \tau, \bmz)
  = \left[~N(\bmx)~N(\bmz)~\right]^{(1-d)/2}~\sum_n e^{s_n\, (t-\tau)}~
    \text{Res}\big[~\tilg_R(\bmx\,; \bmz|s)\,;\, s_n~\big] + \cdots~,
\label{eq:append-inv_L_trans-res}
\end{align}
and the corresponding frequency for each pole $\omega_n = i\, s_n$
($\Im(\omega_n) < \sigma$) is called a quasinormal frequency.
So, a quasinormal frequency is given by a pole of
the retarded Green function in the frequency domain,
$\tilg_R(\bmx\,; \bmz|s)$.

In order to obtain $\tilg_R$,
one must pay attention to the boundary conditions
for Eq.(\ref{eq:tilg_eqn}).
We consider two cases separately.

\begin{enumerate}

\item[(i)] For asymptotically flat black hole spacetimes,
there are two asymptotic regions,
the near-horizon region and spatial infinity.
Equation~(\ref{eq:tilg_eqn}) often has simple forms in these regions:
\begin{align}
  & \big( s^2 - \partial_{r_*}^2 \big)\,
    \tilg_R(\bmx\,; \bmz | s) \sim 0~,
\label{eq:tilg_eqn-AFBH}
\end{align}
where we consider a massless scalar field for simplicity.
The tortoise coordinate of $x$, $r_*$, is defined such that 
the spatial infinity (the horizon) corresponds to
$r_* \rightarrow \infty~(- \infty)$.
There are two independent solutions
$e^{ \pm s\, r_*}$ in each asymptotic region.

Recalling that $\tilg_R$ for $\Re(s) < \sigma$ is given
by the analytic continuation from $\Re(s) > \sigma$,
we first consider the boundary conditions of $\tilg_R$
for $\Re(s) > \sigma$.
Since diverging solutions are unphysical, 
$\tilg_R$ for $\Re(s) > \sigma$ should behave as
\begin{align}
  & \tilg_R(\bmx\,; \bmz|s) \rightarrow
    e^{ s\, r_* } \hspace{0.3cm} (e^{ - s\, r_* })
& & \text{~~for~~} r_* \rightarrow - \infty \hspace{0.3cm} (\infty)~,
\label{eq:tilg_bc-AFBH}
\end{align}
so that
\begin{align}
  & G_R(t, \bmx\,; 0, \bmz)
  \sim e^{s t}~\tilg_R(\bmx\,; \bmz|s)
  \rightarrow e^{ s\, ( t + r_* ) }
    \hspace{0.3cm} (e^{ s\, ( t - r_* ) })
& & \text{~~for~~} r_* \rightarrow - \infty \hspace{0.3cm} (\infty)~.
\label{eq:G_bc-AFBH}
\end{align}
This means that the appropriate boundary conditions 
for the retarded Green function $G_R$
in this case
are a purely ingoing wave near the horizon
and an outgoing wave at the infinity.
Since $\tilg_R$ for $\Re(s) < \sigma$ is defined by
the analytical continuation from $\Re(s) > \sigma$,
$\tilg_R$ for $\Re(s) < \sigma$ also satisfies
the same boundary conditions (\ref{eq:tilg_bc-AFBH}),
in spite of the diverging behavior.

\item[(ii)] For asymptotically AdS black hole spacetimes, 
the tortoise coordinate $\eta_*$ is defined such that
the null infinity (the horizon) corresponds to
$\eta_* \rightarrow 0$ ($-\infty$).
Although Eq.(\ref{eq:tilg_eqn}) has the same form as
Eq.(\ref{eq:tilg_eqn-AFBH}) near the horizon,
Eq.(\ref{eq:tilg_eqn}) has a different form
near the null infinity,
\begin{align}
  & \left( - \partial_{\eta_*}^2 + C/\eta_*^2 \right)
    \tilg_R(\bmx\,; \bmz | s) \sim 0
& & \text{for~~} \eta_* \rightarrow 0~,
\label{eq:tilg_eqn-AdSBH}
\end{align}
where $C$ is a positive constant.
Thus, for $\eta_* \sim 0$, we obtain
$\tilg_R(\bmx\,; \bmz | s) \sim \eta_*^{ 1/2 \pm \sqrt{C + 1/4} }$
or $G_R(t, \bmx\,; 0, \bmz) \sim \eta_*^{ d/2 \pm \sqrt{C + 1/4} }
=: \eta_*^{\Delta_\pm}$, where we use 
$N^{(1-d)/2}(\bmx) \propto \eta_*^{(d-1)/2}$ for $\eta_* \sim 0$.
Since we have $\Delta_- < 0 < \Delta_+$ in many cases, 
we henceforth assume $\Delta_- < 0 < \Delta_+$.

Again, we first consider $\tilg_R$ for $\Re(s) > \sigma$.
Since diverging solutions are physically unacceptable,
$\tilg_R$ for $\Re(s) > \sigma$ should behave as
\begin{align}
  & \tilg_R(\bmx\,; \bmz|s) \rightarrow
  \begin{cases}
    e^{ s\, \eta_* }
      & \text{~~for~~} \eta_* \rightarrow -\infty
  \\
    \eta_*^{ 1/2 + \sqrt{C + 1/4} }
      & \text{~~for~~} \eta_* \rightarrow 0
  \end{cases}~,
\label{eq:tilg_bc-AdSBH}
\end{align}
so that
\begin{align}
  & G_R(t, \bmx\,; 0, \bmz)
  \sim e^{s t}~\tilg_R(\bmx\,; \bmz|s)
  \rightarrow
  \begin{cases}
    e^{ s\, ( t + \eta_* ) }
      & \text{~~for~~} \eta_* \rightarrow -\infty
  \\
    e^{s\, t}~\eta_*^{\Delta_+}
      & \text{~~for~~} \eta_* \rightarrow 0
  \end{cases}~.
\label{eq:G_bc-AdSBH}
\end{align}
We should set the same boundary conditions (\ref{eq:tilg_bc-AdSBH})
to $\tilg_R$ for $\Re(s) < \sigma$
because $\tilg_R$ for $\Re(s) < \sigma$ is defined
by the analytical continuation from $\Re(s) > \sigma$.

\end{enumerate}

To summarize, quasinormal frequencies are obtained by finding
the poles of $\tilg_R$ satisfying Eq.(\ref{eq:tilg_eqn})
with the boundary conditions.
However, it is easy to show that, for a pole of $\tilg_R$, $s_n$,
there exists an eigenmode $\tilphi_n(\bmx)$
of the homogeneous equation of Eq.(\ref{eq:tilg_eqn})
\begin{align}
  & \left( s^2 - \Delta + \cdots \right)\, \tilphi_n(\bmx) = 0~,
\label{eq:tilphi_n-eqn}
\end{align}
which satisfies the boundary conditions (\ref{eq:tilg_bc-AFBH})
for asymptotically flat black hole spacetimes
\cite{ref:KokkotasSchmidt},
and (\ref{eq:tilg_bc-AdSBH})
for asymptotically AdS black hole spacetimes.
Thus, in practice, one can also find the quasinormal frequencies
by solving this eigenvalue problem.
In the main text,
we solve this eigenvalue problem (\ref{eq:tilphi_n-eqn})
with the boundary conditions (\ref{eq:tilg_bc-AdSBH}).
 

\begin{thebibliography}{99}

\bibitem{Witten:1998zw}
E.~Witten,
  Adv.\ Theor.\ Math.\ Phys.\  {\bf 2}, 505 (1998),~
  ``Anti-de Sitter space, thermal phase transition,
  and confinement in  gauge theories,''
  [arXiv:hep-th/9803131].
%
\bibitem{Kraus:2002iv}
P.~Kraus, H.~Ooguri and S.~Shenker, \prd{\bf 67}, 124022 (2003),~
  ``Inside the horizon with AdS/CFT,''
  [arXiv:hep-th/0212277].
%
\bibitem{Hong Liu:2005}
G. Festuccia and Hong Liu, hep-th/0506202,
  ``Excursion beyond the horizon: Black hole singularities 
   in Yang-Mills theories.''
%
\bibitem{Maldacena:2001kr}
J.~M.~Maldacena,
  J.\ High Energy Phys.\ 04 (2003) 021,~
  ``Eternal black holes in Anti-de-Sitter,''
  [arXiv:hep-th/0106112].
%
\bibitem{Hawking:2005kf}
S.~W.~Hawking, hep-th/0507171,~
  ``Information loss in black holes.''
%
\bibitem{GL}
R.~Gregory and R.~Laflamme,
  \prl {\bf 70}, 2837 (1993),~
  ``Black strings and p-branes are unstable,''
  [arXiv:hep-th/9301052];~
%
  Nucl.\ Phys.\ B {\bf 428}, 399 (1994),~
  ``The Instability of charged black strings and p-branes,''
  [arXiv:hep-th/9404071];~
%
  \prd{\bf 51}, R305 (1995),~
  ``Evidence for stability of extremal black p-branes,''
  [arXiv:hep-th/9410050].
%
\bibitem{Aharony:2004ig}
O.~Aharony, J.~Marsano, S.~Minwalla and T.~Wiseman,
  Class.\ Quant.\ Grav.\  {\bf 21}, 5169 (2004),~
  ``Black hole - black string phase transitions
  in thermal 1+1 dimensional supersymmetric Yang-Mills theory
  on a circle,''
  [arXiv:hep-th/0406210].
%
\bibitem{Hawking:1982dh}
S.~W.~Hawking and D.~N.~Page,
  Commun.\ Math.\ Phys.\  {\bf 87}, 577 (1983),~
  ``Thermodynamics Of Black Holes In Anti-De Sitter Space.''
%
\bibitem{Klebanov:2000hb}
I.~R.~Klebanov and M.~J.~Strassler,
  J.\ High Energy Phys.\ 08 (2000) 052,~
  ``Supergravity and a confining gauge theory: Duality cascades and
  $\chi$SB-resolution of naked singularities,''
  [arXiv:hep-th/0007191].
%
\bibitem{Polchinski:2000uf}
J.~Polchinski and M.~J.~Strassler, hep-th/0003136,~
  ``The string dual of a confining four-dimensional gauge theory.''
%
\bibitem{Itzhaki:1998dd}
N.~Itzhaki, J.~M.~Maldacena, J.~Sonnenschein and S.~Yankielowicz,
  \prd{\bf 58}, 046004 (1998),~
  ``Supergravity and the large N limit of theories with sixteen
  supercharges,''
  [arXiv:hep-th/9802042].
%
\bibitem{Brandhuber:1998er}
A.~Brandhuber, N.~Itzhaki, J.~Sonnenschein and S.~Yankielowicz,
  J.\ High Energy Phys.\ 06 (1998) 001,~
  ``Wilson loops, confinement, and phase transitions
    in large N gauge  theories from supergravity,''
  [arXiv:hep-th/9803263].

\bibitem{Sekino:1999av}
Y.~Sekino and T.~Yoneya,
  Nucl.\ Phys.\ B {\bf 570}, 174 (2000),~
  ``Generalized AdS-CFT correspondence for matrix theory
    in the large N limit,''
  [arXiv:hep-th/9907029].
%

\bibitem{HoHu}
G. T. Horowitz and V. E. Hubeny,
  \prd{\bf 62}, 024027 (2000),~
  ``Quasinormal modes of AdS black holes
    and the approach to thermal equilibrium,''
  [arXiv:hep-th/9909056].

\bibitem{MossNorman}
I. G. Moss and J. P. Norman,
  Class.\ Quant.\ Grav.\ {\bf 19}, 2323 (2002),~
  ``Gravitational quasinormal modes for Anti-de Sitter black holes,''
  [arXiv:gr-qc/0201016].

\bibitem{BK}
E. Berti, K. D. Kokkotas,
  \prd{\bf 67}, 064020 (2003),~
``Quasinormal modes of Reissner-Nordstr\"om-anti-de Sitter black holes:
scalar, electromagnetic and gravitational perturbations'', 
[arXiv:gr-qc/0301052].

\bibitem{CKL}
V. Cardoso, R. Konoplya, and J. P. S. Lemos, 
  \prd{\bf 68}, 044024 (2003),~
  ``Quasi-normal frequencies of Schwarzschild black holes
    in anti-de Sitter spacetime: A complete study of
    the overtome asymptotic behavior,'' 
  [arXiv:gr-qc/0305037].

%
\bibitem{Konoplya}
R. A. Konoplya,
  \prd{\bf 68}, 124017 (2003),~
  ``Gravitational quasinormal radiation
    of higher-dimensional black holes,''
  [arXiv:hep-th/0309030].
%
\bibitem{ref:MusiriSiopsis}
S. Musiri and G. Siopsis,
  Phys.\ Lett.\ B {\bf 563}, 102 (2003),~
  \lq\lq Quasinormal modes of large AdS black holes\rq\rq,
  [arXiv:hep-th/0301081];~
%
  Phys.\ Lett.\ B {\bf 576}, 309 (2003),~
  \lq\lq Asymptotic form of quasi-normal modes
  of large AdS black holes\rq\rq,
  [arXiv:hep-th/0308196].
%
%
\bibitem{Benincasa:2005iv}
  P.~Benincasa, A.~Buchel and A.~O.~Starinets, hep-th/0507026,~
``Sound waves in strongly coupled non-conformal gauge theory plasma,''
  [arXiv:hep-th/0507026].

\bibitem{Policastro:2002se}
G.~Policastro, D.~T.~Son and A.~O.~Starinets,
  J.\ High Energy Phys.\ 09 (2002) 043,~
  ``From AdS/CFT correspondence to hydrodynamics,''
  [arXiv:hep-th/0205052].

\bibitem{ref:Starinets}
A.O. Starinets,
  \prd{\bf 66}, 124013 (2002),~
  \lq\lq Quasinormal modes of near extremal black branes\rq\rq,
  [arXiv:hep-th/0207133].
%
\bibitem{ref:NunezStarinets}
A. N\'{u}\~{n}ez and A.O. Starinets,
  \prd{\bf 67}, 124013 (2003),~
  \lq\lq AdS/CFT correspondence, quasinormal modes,
   and thermal correlators
   in $\cal{N}=4$ supersymmetric Yang-Mills theory\rq\rq,
  [arXiv:hep-th/0302026].
%
\bibitem{ref:KovtunStarinets}
P.K. Kovtun and A.O. Starinets, hep-th/0506184,~
  \lq\lq Quasinormal modes and holography.\rq\rq
%
\bibitem{Polchinski:1998rr}
J.~Polchinski,
  {\it String theory. Vol. 2: Superstring theory and beyond},
   (Cambridge University Press, Cambridge, 1998).

\bibitem{Townsend}
H. J. Boonstra, K. Skenderis, P. K. Townsend, 
  J.\ High Energy Phys.\ 01 (1999) 003,~
  ``The domain-wall/QFT correspondence,''
  [arXiv:hep-th/9807137].

%
\bibitem{ref:KodamaSasaki}
H. Kodama and M. Sasaki,
  Prog.\ Theor.\ Phys.\ Suppl.\ {\bf 78}, 1 (1984),~
  \lq\lq Cosmological Perturbation Theory.\rq\rq
%

\bibitem{Leaver:1990}
E. W. Leaver, 
  \prd{\bf 41}, 2986 (1990),~
  ``Quasinormal modes of Reissner-Nordstr\"om black holes.''
%
\bibitem{ref:NatarioSchiappa}
V. Cardoso, J. Natario, and R. Schiappa,
  J.\ Math.\ Phys.\ {\bf 45}, 4698 (2004),~
  \lq\lq Asymptotic Quasinormal Frequencies for Black Holes
  in Non-Asymptotically Flat Spacetimes,\rq\rq
  [arXiv:hep-th/0403132];~
%
J. Natario and R. Schiappa, hep-th/0411267,~
  \lq\lq On the Classification of Asymptotic Quasinormal Frequencies
  for $d$-Dimensional Black Holes and Quantum Gravity.\rq\rq
%
\bibitem{Iizuka:2003ad}
N.~Iizuka, D.~Kabat, G.~Lifschytz and D.~A.~Lowe,
  \prd{\bf 68}, 084021 (2003),~
  ``Stretched horizons, quasiparticles and quasinormal modes,''
  [arXiv:hep-th/0306209].
%
\bibitem{ref:KokkotasSchmidt}
K.D. Kokkotas and B.G. Schmidt,
  \lq\lq Quasi-Normal Modes of Stars and Black Holes,\rq\rq
  (August, 1999),
  [Article in Online Journal Living Reviews in Relativity]:
   http://relativity.livingreviews.org/Articles/lrr-1999-2/index.html.
%
\bibitem{ref:MorseFeshbach}
P.M. Morse and H. Feshbach,
{\it Methods of Theoretical Physics}, (McGraw-Hill. New York, 1953).
\end{thebibliography}
\end{document}